\documentclass[11pt,twoside]{article}
\usepackage{asp2014}

\aspSuppressVolSlug
\resetcounters

\begin{document}

\title{GMRT Archive Processing Project}

\author{Shubhankar~Deshpande$^1$, Yogesh Wadadekar$^2$, Huib Intema$^3$, B. Ratnakumar$^2$, Lijo George$^2$, Rathin Desai$^2$, Archit Sakhadeo$^2$, Shadab Shaikh$^2$, C. H. Ishwara-Chandra$^2$ and Divya Oberoi$^2$
  \affil{$^1$School of Computer Science, Carnegie Mellon University, 5000 Forbes Avenue, Pittsburgh, PA 15213, USA \email{shubhand@cs.cmu.edu}}
  \affil{$^2$National Centre for Radio Astrophysics, TIFR, Post Bag 3, Ganeshkhind, Pune 411007, India}
\affil{$^3$Leiden Observatory, Leiden University, Niels Bohrweg 2, 2333 CA, Leiden, The Netherlands}
}           

\paperauthor{Shubhankar~Deshpande}{shubhand@cs.cmu.edu}{ORCID}{Carnegie Mellon University}{Author1 Department}{City}{State/Province}{Postal Code}{Country}
\paperauthor{Sample~Author2}{Author2Email@email.edu}{ORCID_Or_Blank}{Author2 Institution}{Author2 Department}{City}{State/Province}{Postal Code}{Country}


\begin{abstract}

The GMRT Online Archive now houses over 120 terabytes of
interferometric observations obtained with the GMRT since the
observatory began operating as a facility in 2002. The utility of this
vast data archive, likely the largest of any Indian telescope, can be
significantly enhanced if first look (and where possible, science
ready) processed images can be made available to the user
community. We have initiated a project to pipeline process GMRT images
in the 150, 240, 325 and 610 MHz bands. The thousands of
processed continuum images that we will produce will prove useful in
studies of distant galaxy clusters, radio AGN, as well as nearby
galaxies and star-forming regions. Besides the scientific returns, a
uniform data processing pipeline run on a large volume of data can be
used in other interesting ways. For example, we will be able to
measure various performance characteristics of the GMRT telescope and
their dependence on waveband, time of day, RFI environment, backend,
galactic latitude etc. in a systematic way. A variety
of data products such as calibrated UVFITS data, sky images and AIPS
processing logs will be delivered to users via a web-based
interface. Data products will be compatible with standard Virtual
Observatory protocols.
  
\end{abstract}

\section{Introduction}

The Giant Meterwave Radio Telescope \citep[GMRT,][]{swarup91} is a low frequency radio
interferometer operating at a site 80 km north of Pune, India. Since
2002, it has been operated as an international open access facility by
India's National Centre for Radio Astrophysics. All interferometric observations carried out
with the GMRT have been carefully archived over the years using
several different tape and disk based storage technologies (see Fig.\ref{gmrtsky}). Raw interferometric visibilities were made available to the international user community, on request, via DVDs until 2009. Thereafter, all raw data
were made accessible for search and download via a password authenticated,
web-based interface, the NCRA Archive and Proposal handling System
(NAPS)\footnote{http://naps.ncra.tifr.res.in/goa}

\articlefigure{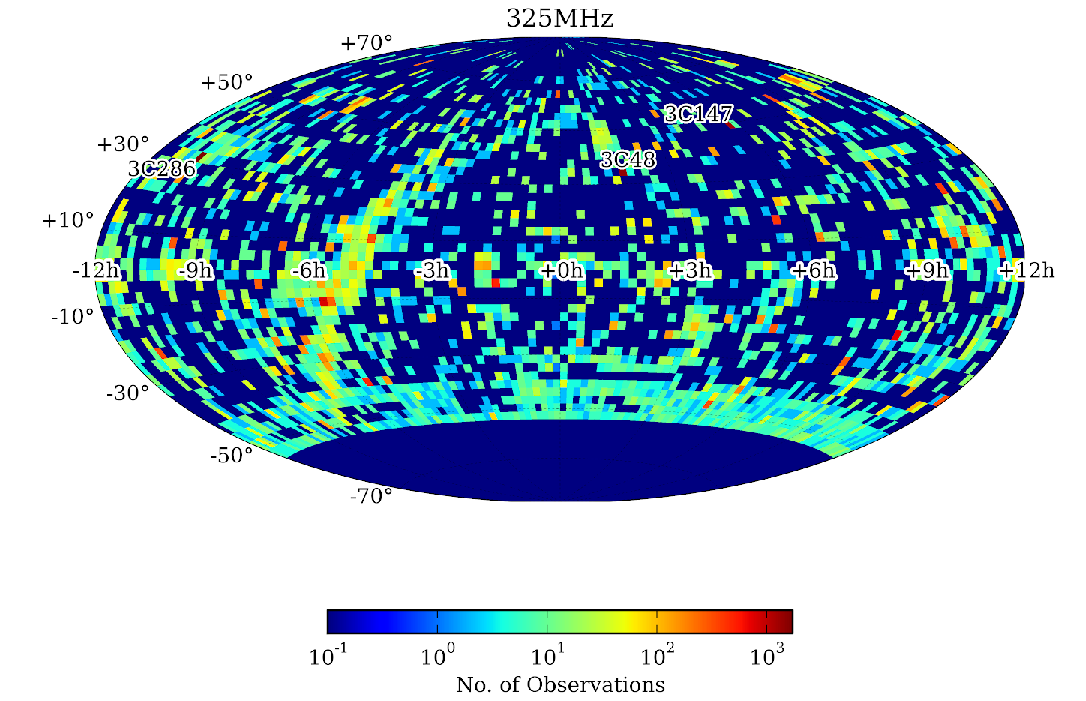}{gmrtsky}{Observations made with the GMRT in equatorial coordinates at 325 MHz, with the sky pixellated into a 3x3 degrees grid for each datapoint. Much of the sky north of the declination limit of the GMRT has been covered. The 3 primary flux calibrators - 3C 48, 3C 147 and 3C 286 have been observed on numerous occasions. Some part of the galactic plane has also been extensively observed. These raw data will be processed into images by our project.}

The NAPS system now hosts over 120 TB of data and delivers them to the
GMRT user community located in about 40 countries worldwide. In 2018,
we saw an average of about 55 data requests per month with an average
size of 50 GB per request. Despite this high level of usage of the
archive, the scientific utility of these data is greatly limited by
the large effort required to transform these raw visibilities into
science ready images. To address this situation, we have initiated an effort
to generate pipeline processed continuum images for GMRT data.  We
are looking to provide users with ``first look'' (worst case) and
``science ready'' (best case) images for as many GMRT observations as
possible. A ``certifiably bad'' tag on data is also useful, because it
helps convince the time allocation committee that fresh observations
are warranted.

\section{Imaging the GMRT Archive}

Presently, there is no standard or official pipeline for processing
GMRT data. However, there are several scripts and pipelines developed
by different users, some of which are publicly available. One of the
most sophisticated, publicly available, pipelines for processing data
from the GMRT is the Source Peeling and Modelling (SPAM) pipeline
developed by H. Intema \citep{intema17}. It was used to successfully
pipeline process about 2000 hours of GMRT data from the TIFR GMRT Sky
Survey \citep[TGSS ADR1\footnote{http://tgssadr.strw.leidenuniv.nl},][]{intema17}. SPAM is a Python module that provides an interface
to AIPS via ParselTongue \citep{kettenis06} and ObitTalk
\citep{cotton08}. ParselTongue provides access to AIPS tasks, data
files (images \& visibilities) and tables. SPAM extensively uses
several other popular Python modules like numpy and scipy. Data
reductions are carried out by well-tested Python scripts that executes
AIPS tasks directly or via high-level functions that make multiple
AIPS or ParselTongue calls. SPAM now also includes a fully automated
pipeline for reducing legacy GMRT observations at 150, 235, 325 and
610 MHz.
  
\section{Building our compute infrastructure}

We used a simple Beowolf cluster architecture for our processing. A
master node acts as a fileserver for the compute stack - AIPS, SPAM,
Obit and Parseltongue  which is NFS exported to a set of compute nodes
(simple headless desktop computers) which have a standard Ubuntu 16.04
server installation plus some additional software libraries installed
via Tentakel from the master node. Ganglia was chosen as the tool for
monitoring cluster status. After some experimentation, we found that
using a Docker container to install our software stack on each compute
node was more efficient. We began processing with a four node cluster
which was gradually expanded to include about 30 headless
desktops. Even with this modest hardware, it is possible to process
about 5 months of GMRT data in about a month. The raw data archive is
hosted on a Dell EMC Isilon system from where it is NFS exported to
our data processing cluster. After processing, the outputs are copied
back onto the Isilon system for long term storage and disaster
recovery compliant backup.

\section{Data processing and delivery}

The SPAM pipeline is designed to be used interactively by a single
user and each processing instance runs as a single thread on the
CPU. We wrote a set of Python and bash scripts to make SPAM operate in
non-interactive fashion and to run multiple processing threads simultaneously
on each multicore computer in our cluster.

We realised quickly that keeping track of the processing was very
cumbersome since data processing rarely progressed linearly. Failures
could happen due to poor data quality or due to some limitation in
SPAM. It was important to bookkeep all of these so that we could get
an accurate picture of the current status of the processing for each
observation and to gather statistics on failure situations. We have
developed a comprehensive database schema to keep track of the
processing. Scattered throughout the SPAM processing are read and
write calls to the database recording successes and failure and
metadata on them. This database is also critical in determining which
outputs are ready to be delivered to users after some automated and
manual quality control. The database will also prove useful in
analysis of the long term trends at the observatory in terms of the
evolving telescope charateristics with manmade radio frequency
interference and the ionospheric environment.

For datasets where the processing is successful (see
Fig.\ref{sampleimages}) a variety of data products such as calibrated
UVFITS data, sky images, AIPS processing logs are generated and are
now being integrated into the NAPS system for delivery. These will
become visible to NAPS users as additional value added data products
which will be compatible with standard Virtual Observatory protocols.

\articlefiguretwo{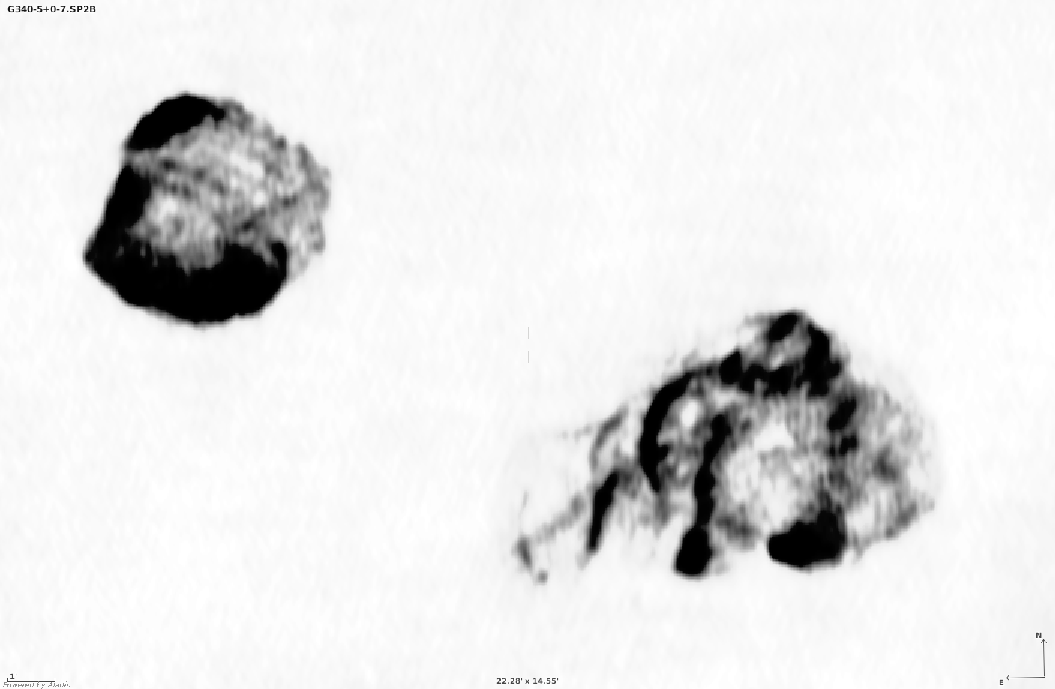}{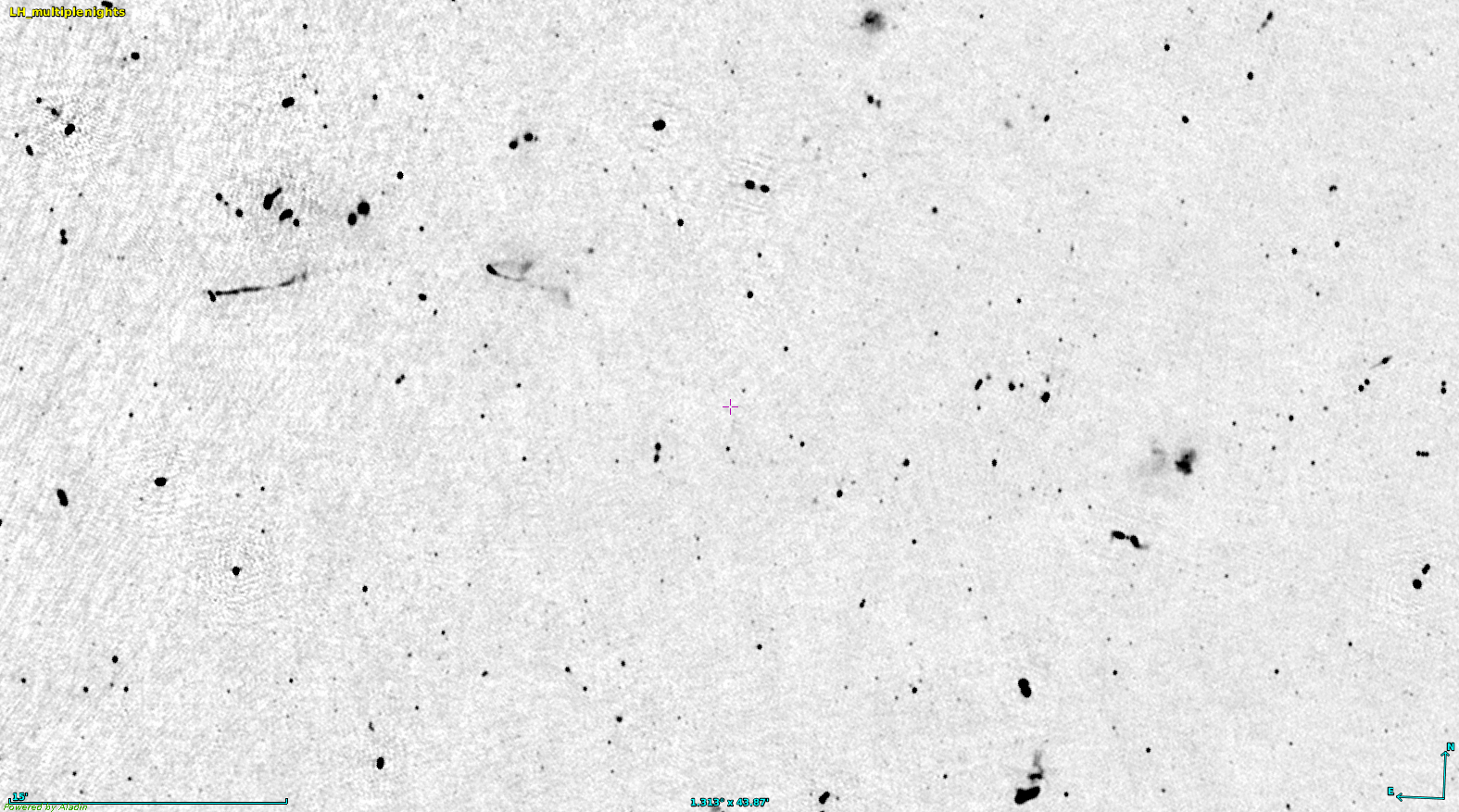}{sampleimages}{{\emph Left:} Two supernova remnants near the galaxy centre observed at 325 MHz with the GMRT. {\emph Right:} Portion of a 325 MHz image of the Lockman Hole field. The extended filamentary souces at top left are cluster radio relics. $>5000$ radio sources are seen over $\sim12$ deg$^2$. These are just two of the thousands of radio images being produced by the GMRT Archive Processing Project.}

\section{Future plans}

The current SPAM pipeline only works on legacy GMRT data. For data which is now
streaming from the upgraded GMRT \citep[uGMRT,][]{gupta17}, with
 seamless frequency coverage and large bandwidth, a different
pipeline would be needed. It would also be tremendously useful if the
data processing can be done in near real time, so that authorised
users can view the processed images from their own observations,
within a few days. We are currently exploring the software and hardware
enhancements that are necessary to enable these exciting possibilities
over the next few years.



\bibliography{P4-7}

\end{document}